\pdfoutput=1
\documentclass{article}

\usepackage{arxiv}

\usepackage[utf8]{inputenc} 
\usepackage[T1]{fontenc}    
\usepackage{hyperref}       
\usepackage{url}            
\usepackage{booktabs}       
\usepackage{amsfonts}       
\usepackage{amsmath, amssymb} 
\usepackage{nicefrac}       
\usepackage{microtype}      
\usepackage{graphicx}
\usepackage{natbib}
\usepackage{doi}
\usepackage{tabularx}       

\title{Bridging Number-Theoretic Primality Testing and Algorithmic Optimization: An Adaptive Cryptographic Framework}

\date{}

\author{
    Tran Minh Thanh\textsuperscript{a,b}, \ Vo Hoai Viet\textsuperscript{a,b,*} \\[1.5ex]
    \normalfont\small\textsuperscript{a}\textit{Faculty of Information Technology, University of Science, Ho Chi Minh City, Vietnam} \\
    \normalfont\small\textsuperscript{b}\textit{Vietnam National University, Ho Chi Minh City, Vietnam}
}

\renewcommand{\headeright}{}
\renewcommand{\undertitle}{}
\renewcommand{\shorttitle}{\textit{Adaptive Primality Testing Framework}}

\hypersetup{
pdftitle={Bridging Number-Theoretic Primality Testing and Algorithmic Optimization: An Adaptive Cryptographic Framework},
pdfsubject={cs.CR, math.NT},
pdfauthor={Minh Thanh Tran, Vo Hoai Viet},
pdfkeywords={Primality Testing, Arbitrary-Precision Arithmetic, Machine Learning, Elliptic Curve Cryptography, Optimization Paradigms, Information Security, Quantum Entropy, QRNG.},
}

\begin{document}
\maketitle
\begingroup
\renewcommand{\thefootnote}{}
\footnotetext{\textsuperscript{*}Corresponding author: \texttt{vhviet@fit.hcmus.edu.vn}}
\endgroup

\begin{abstract}
The security of modern public-key cryptographic systems, such as RSA, Diffie-Hellman, and ECC, depends entirely on the computational intractability of the integer factorization and discrete logarithm problems. For these asymmetric algorithms to operate, a core requirement is that the system must be capable of generating, testing, and simultaneously proving the primality of extraordinarily large numbers. To meet those requirements, this study proposes a comprehensive framework, bridging classical number theory with modern machine learning models and optimization algorithms. Specifically, the paper delineates an adaptive, arbitrary-precision architecture dedicated exclusively to primality testing. Concurrently, the research deeply analyzes the theoretical foundations of testing methods (both deterministic and probabilistic), hardware-level optimization techniques, and the application of quantum-secured entropy sources, while providing rigorous mathematical proofs to guarantee security even when the prime distribution is biased.
\end{abstract}

\keywords{Primality Testing \and Arbitrary-Precision Arithmetic \and Machine Learning \and Elliptic Curve Cryptography \and Optimization Paradigms \and Information Security \and Quantum Entropy \and QRNG.}

\section{Introduction}

\subsection{Mathematical Foundations of Prime Numbers}
The Fundamental Theorem of Arithmetic indicates that every positive integer greater than one can be uniquely factorized into a product of prime numbers, independent of their ordering. By definition, prime numbers are simply those that possess exactly two distinct divisors: one and the number itself. Although this theoretical foundation is simple and has been studied since around 300 B.C. in Hellenistic Alexandria, distinguishing between a prime number and a composite number becomes a formidable challenge when the numbers reach extremely large scales.

\subsection{Trial Division and Complexity Limits}
To test the primality of a number $N$, the most 'naive' method commonly cited is Trial Division. This algorithm operates on an exhaustive search principle, sequentially attempting to divide $N$ by all integers less than or equal to $\sqrt{N}$. A number is classified as a composite if an exact division occurs; conversely, it is proven to be prime if it passes the entire test. However, the programmatic simplicity and absolute accuracy of this algorithm are accompanied by a formidable challenge: its arithmetic complexity reaches $O(\sqrt{N})$ divisions, which translates to a prohibitive bit complexity of $O(\sqrt{N} \log^2 N)$ in arbitrary-precision environments. This characteristic causes the execution time to grow exponentially based on the number of input bits. Consequently, when confronted with a 1000-digit integer, Trial Division is rendered entirely useless due to exceeding the processing limits of the system.

\begin{figure}[htbp]
    \centering
    \includegraphics[width=0.77\textwidth]{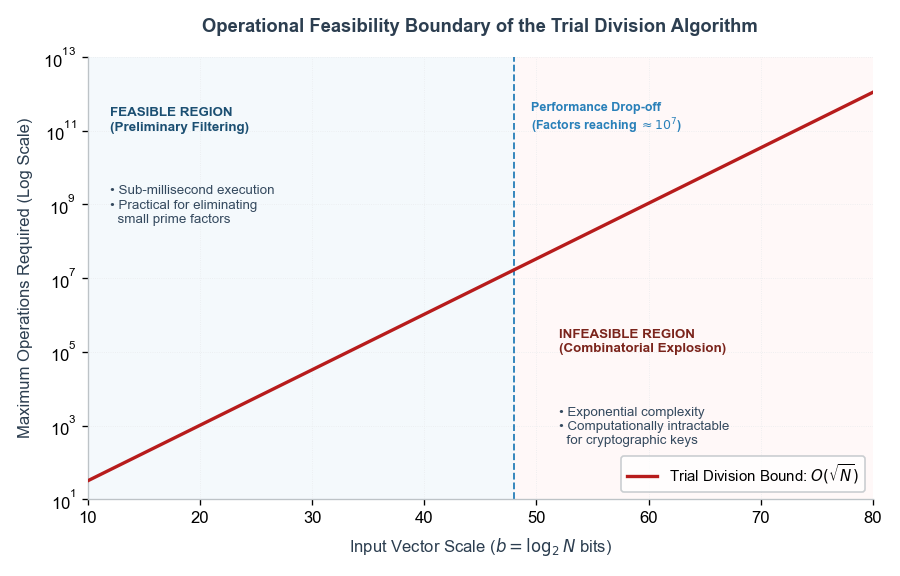}
    \caption{Complexity explosion and feasibility limits of the trial division algorithm.}
    \label{fig:trial_division}
\end{figure}

\subsection{Cryptographic Bottlenecks and Motivations}
For centuries, resolving composite numbers into their prime factors was primarily of academic interest. Carl Friedrich Gauss once asserted this to be one of the most important and useful problems in arithmetic, deserving a high position in science. However, the explosion of secure digital communications in the late 20th century transformed primality testing from a purely theoretical pursuit into a rigorously constrained engineering problem. As cryptographic key sizes continuously expand from 1024-bit to 4096-bit and beyond, in order to resist modern cryptanalytic techniques and threats from quantum computers, traditional primality testing algorithms begin to reveal severe performance weaknesses.

To generate a secure prime number, the system must perform hundreds of probabilistic tests. This process requires extremely computationally expensive modular exponentiations on integers whose sizes far exceed the processing limit (word size) of conventional Arithmetic Logic Units (ALUs). Consequently, modern cryptographic software is compelled to rely on arbitrary-precision arithmetic architectures, alongside deeply optimized finite field reductions and meticulous search models to balance computational costs with mathematical reliability~\cite{ref3}.  

Furthermore, the integration of advanced optimization models—ranging from continuous gradient descent and discrete policy gradients to Bayesian optimization and supervised machine learning—has contributed to opening a new era for algorithmic number theory. Although classical machine learning models and continuous gradient descent fail against the high-frequency periodicity of modular arithmetic~\cite{ref7}, hybrid approaches can demonstrate outstanding effectiveness. They provide promising pathways to optimize the search space, dynamically fine-tune sieve block sizes to hardware memory limits, and evaluate exploratory heuristics for intelligent "witness" selection in probabilistic tests to minimize modular exponentiation overhead.

Based on this context, this research proposes a comprehensive framework acting as a bridge between classical number theory and contemporary machine learning and optimization models. The paper delineates a flexible, holistic architecture capable of arbitrary-precision computation, specifically tailored for the primality testing problem in cryptography. This framework not only evaluates the theoretical foundations of deterministic and probabilistic tests but also delves into hardware optimization techniques as well as the relationship between integer factorization and primality proving. At the same time, the research presents the implementation of heuristic search algorithms, accompanied by rigorous mathematical proofs to ensure absolute security even when the prime distribution is biased.

\section{Mathematical Preliminaries and the Architecture of Arbitrary Precision}

While the concept of prime numbers is simple, their actual distribution does not follow a predictable pattern. Their asymptotic density is described by the Prime Number Theorem, where as $x$ approaches infinity, the number of primes less than or equal to $x$, denoted as $\pi(x)$, is approximately $x/\ln(x)$~\cite{ref_math_textbook}.

When computing at cryptographic scales, systems must address hardware constraints. Standard Arithmetic Logic Units (ALUs) only process data in fixed 32-bit or 64-bit chunks. Meanwhile, cryptographic algorithms always utilize colossal integers spanning hundreds to thousands of decimal digits, equivalent to 1024, 2048, or 4096 bits. Thus, the computational foundation of any primality testing framework is the ability to perform arbitrary-precision arithmetic, or multiple-precision arithmetic~\cite{ref2}.

In this environment, large integers are dynamically allocated as variable-length arrays of machine-word digits. Popular libraries such as the GNU Multiple Precision Arithmetic Library (GMP), FLINT, and the Rust-based Malachite help translate theoretical algorithms into safe and hardware-optimized execution architectures~\cite{ref8}. These implementations ensure that addition and subtraction execute digit-by-digit with $O(\log N)$ complexity, whereas classical multiplication and division take up to $O((\log N)^2)$ time~\cite{ref2}.  

The primary computational bottleneck in evaluating primality—typically when calculating the modular exponentiation sequence $a^d, a^{2d}, a^{4d} \pmod N$ in modern testing algorithms—is the reduction operation (modular reduction). If standard modular multiplication is used, the system must repeatedly perform highly expensive trial divisions to find the remainder. High optimization of this discrete operation is a mandatory requirement for a cryptographic framework to achieve practical feasibility.  

To thoroughly resolve this difficulty, Peter L. Montgomery (1985) introduced Montgomery modular multiplication. This algorithm completely bypasses the need for multi-precision division by shifting computations to a dedicated arithmetic space known as the Montgomery domain. Given a large odd integer $N$ (the modulus) and an auxiliary radix $R$ (with $R > N$ and $\gcd(R,N)=1$), values are mapped into this domain. The radix $R$ is canonically chosen as a power of 2 matching the ALU word size, such as $2^{1024}$. A standard integer $a$ is transformed into its corresponding Montgomery representation~\cite{ref9}:  $$\tilde{a} = aR \pmod N$$
The Montgomery multiplication of two mapped values $\tilde{a}$ and $\tilde{b}$ yields the following result:  
$$Mont(\tilde{a}, \tilde{b}) = \tilde{a}\tilde{b}R^{-1} \pmod N$$
The utility of this algorithm lies in its optimal execution at the hardware level. Because $R$ is chosen as a power of 2, the modulo $R$ operation and division by $R$ are entirely replaced by inexpensive binary bit-shifts and additions, completely eliminating the costly trial-division logic that hampers computational speed. Although converting data into and out of the Montgomery domain incurs a certain setup overhead, modular exponentiations require hundreds of sequential multiplications that take place almost exclusively within this domain, rendering the initial and final transformation costs modest compared to the overall computational cost~\cite{ref9}.  

Alongside Montgomery arithmetic, Barrett reduction is an effective alternative choice that optimizes modular arithmetic by estimating quotients based on precomputed invariants. For a specific modulus $N$ of $k$ words (or bits) in base $b$, Barrett reduction precomputes the reciprocal invariant~\cite{ref10}:  
$$\mu = \lfloor b^{2k} / N \rfloor$$
To compute $a \pmod N$ for an operand $a < b^{2k}$, the algorithm estimates the quotient $q \approx \lfloor a/N \rfloor$ by multiplying only the most significant part of $a$ with $\mu$ and performing word shifts~\cite{ref2, ref10}:  
$$q = \lfloor (\lfloor a / b^{k-1} \rfloor \cdot \mu) / b^{k+1} \rfloor$$
The remainder is subsequently extracted via $r = a - qN$, requiring at most two conditional subtractions of $N$ to guarantee $0 \le r < N$~\cite{ref2, ref10}.

Barrett reduction is particularly effective in cryptographic contexts where the modulus $N$ changes infrequently, or when transforming numbers into the Montgomery representation creates an unnecessary bottleneck for short-lived computations. Consequently, modern cryptographic systems today often use both algorithms—Montgomery and Barrett—interchangeably or in parallel, depending on the characteristics of the execution pipeline and the instruction set of the host hardware~\cite{ref10}.

\section{Taxonomy of Primality Testing Frameworks}

\subsection{Deterministic and Polynomial-Time Tests}

The problem of distinguishing prime numbers from composites can be approached through varied algorithmic philosophies. Primality testing algorithms are systematically classified into four primary categories: deterministic tests, Monte-Carlo randomized tests, Las-Vegas randomized tests, and heuristic combinations. Deterministic tests yield an absolute, mathematically exact proof regarding primality, operating without the possibility of error (neither false positives nor false negatives). However, the price paid for this absolute certainty is a prohibitive computational cost for cryptographic-scale integers, necessitating a rigorous evaluation of their practical utility.

The most fundamental deterministic method is Trial Division. Despite its fundamentally exponential time complexity and complete unsuitability for proving large cryptographic primes, trial division still plays a crucial role in modern systems as a preliminary filter to rapidly eliminate candidates with small prime factors (e.g., factors up to $10^7$) before advancing to exponentially faster probabilistic algorithms~\cite{ref1}. Implementing this method at the system level requires careful attention to micro-optimizations. Naive implementations frequently utilize floating-point square root functions $\sqrt{N}$ to establish the loop termination boundary. However, as $N$ approaches the limits of 64-bit unsigned integers, the loss of floating-point precision can cause premature or incorrect loop termination. Evaluating the condition $i \times i \le N$ is similarly hazardous due to severe integer overflow vulnerabilities. The most optimal and secure remedy is replacing it with the strictly bounded integer division condition $i \le N/i$.

Beyond brute-force exhaustion, specific deterministic algorithms exploit the distinct structural properties of specialized integer families. Pepin's Test (1877) is a prime example, engineered specifically for Fermat numbers of the form $F_n = 2^{2^n} + 1$. Pepin's theorem states that $F_n$ is prime if and only if it satisfies the congruence:

$$3^{(F_n - 1)/2} \equiv -1 \pmod{F_n}$$
Executing in $O(\log^2 N)$ time, this algorithm operates highly efficiently and yields a definitive result. Nevertheless, because the base is fixed and the structural prerequisite is absolute, Pepin's test cannot be applied to other integer forms (yielding false negatives for standard prime numbers), rendering it mathematically incompatible with general cryptographic key generation~\cite{ref1}.

Similarly, the Lucas-Lehmer Test is constructed exclusively for Mersenne numbers of the form $M_p = 2^p - 1$ (where $p$ is an odd prime). The test establishes a sequence with an initial value $s_0 = 4$, generating subsequent terms via the recurrence relation:
$$s_i = (s_{i-1}^2 - 2) \pmod{M_p}$$
The theorem states that $M_p$ is prime if and only if the term $s_{p-2} = 0$. Operating with a complexity of $O(\log^2 N \log \log N)$, Lucas-Lehmer is the benchmark algorithm deployed in distributed computing projects (such as GIMPS) to discover record-breaking massive primes. However, much like Pepin's test, its applicability is strictly confined to the Mersenne form, rendering it useless for verifying randomly generated integers~\cite{ref1}.

For decades, the pursuit of a generalized, unconditional deterministic primality test executing in polynomial time was considered the "holy grail" of algorithmic number theory. This was definitively resolved in August 2002 when Manindra Agrawal, Neeraj Kayal, and Nitin Saxena published the AKS algorithm, formally proving that "PRIMES is in P". The AKS test generalizes Fermat's Little Theorem to polynomials with coefficients in the ring $\mathbb{Z}/n\mathbb{Z}$. Although subsequent research optimized the theoretical complexity of AKS from $O(\log^{11.5} N)$ down to $\tilde{O}(\log^6 N)$, in practice the algorithm suffers from massive constant overheads. The required complex polynomial expansions render AKS drastically slower than probabilistic methods when handling small- to medium-sized cryptographic integers, meaning it is rarely applied in real-world cryptographic key generation pipelines~\cite{ref1}.

\subsection{Probabilistic Tests and Monte-Carlo Algorithms}

Because deterministic primality tests like AKS have cryptographic key generation speeds that are too slow in real time, practical systems primarily rely on Monte-Carlo randomized tests. Fundamentally, these algorithms operate as filters to discover composite numbers. When identifying a number as a composite, the provided result is 100\% accurate; but when they conclude a number is a prime, that number is only considered a "probable prime" accompanied by a margin of error that can be strictly calculated~\cite{ref1}.

The simplest algorithm in this group is the Fermat Primality Test, built upon Fermat's Little Theorem. This theorem proves that if $p$ is a prime number, then for every integer $a$ satisfying $1 < a < p-1$, the following congruence is always true:
$$a^{p-1} \equiv 1 \pmod p$$
To test the integer $N$, the algorithm will randomly choose a base $a$ and calculate the congruence:
$$a^{N-1} \equiv 1 \pmod N$$
through arbitrary-precision modular exponentiation. If the calculation is not satisfied, $a$ is considered a "Fermat witness" and $N$ is definitively a composite number. Conversely, if the calculation is satisfied, $N$ is likely a prime number. However, if $N$ is actually an odd composite number but $a$ still satisfies the equation, then $a$ is called a "Fermat liar". The fatal weakness of the Fermat method lies in the existence of Carmichael numbers, which are composite numbers (such as 561) that always pass the Fermat condition with every base coprime to them. Because the quantity of Carmichael numbers is infinite, the failure rate of the Fermat algorithm against these composite numbers is 100\%, making the algorithm insufficiently secure to be used independently in cryptographic systems~\cite{ref1}.

Introduced in 1977, the Solovay-Strassen Primality Test overcame the disadvantages related to Carmichael numbers by using the Euler criterion combined with the Legendre and Jacobi symbols. This theorem states that if $p$ is a prime number then:
$$a^{(p-1)/2} \equiv (a/p) \pmod p$$
where $(a/p)$ represents the Legendre symbol. To test the primality of $N$, distinguished values for $a$ are selected and the above congruence is verified. If any base invalidates the equation, $N$ is concluded to be a composite number. A prominent advantage of the Solovay-Strassen algorithm is that no composite number can "deceive" this algorithm with every base; thereby, Carmichael numbers are successfully identified as composite numbers. However, the algorithm can still miss "Euler pseudoprimes" (odd composite numbers that satisfy the test with certain bases). In addition, having to calculate the Jacobi symbol creates a massive processing burden, which has significantly reduced execution speed compared to more modern models~\cite{ref1}.

The Miller-Rabin Primality Test, initially proposed as a deterministic algorithm based on the unproven Extended Riemann Hypothesis and later transformed by Michael Rabin into an unconditional probabilistic algorithm, currently plays the role as the core of modern cryptography~\cite{ref14}. This algorithm developed from the logic of Fermat's theorem by exploiting the mathematical properties of non-trivial square roots of 1 modulo $N$. With an odd integer $N > 2$, the value $N-1$ is always an even number and can be uniquely factorized into the form $2^s \cdot d$, where $d$ is an odd number. When randomly choosing a base $a$, the algorithm will calculate the sequence of modular exponentiations $a^d, a^{2d}, a^{4d}, \dots, a^{2^{s-1}d} \pmod N$. The number $N$ will be recognized as a "strong probable prime" if it satisfies:
$$a^d \equiv 1 \pmod N \quad \text{or} \quad a^{2^j \cdot d} \equiv -1 \pmod N$$
with any value $0 \le j < s$~\cite{ref1}.

The strictness of the Miller-Rabin test is extremely impressive mathematically. The probability that a composite number can pass a single loop of the test (acting as a strong pseudoprime) is always less than $1/4$. If we repeat the test with $k$ independent random bases, the error rate of misidentifying a composite number as a prime number will drop deeply to only $(1/4)^k$~\cite{ref1}. Operating with a polynomial time complexity of $O(k \log^3 N)$, this algorithm provides extremely efficient computational performance.

\begin{figure}[htbp]
    \centering
    \includegraphics[width=0.83\textwidth]{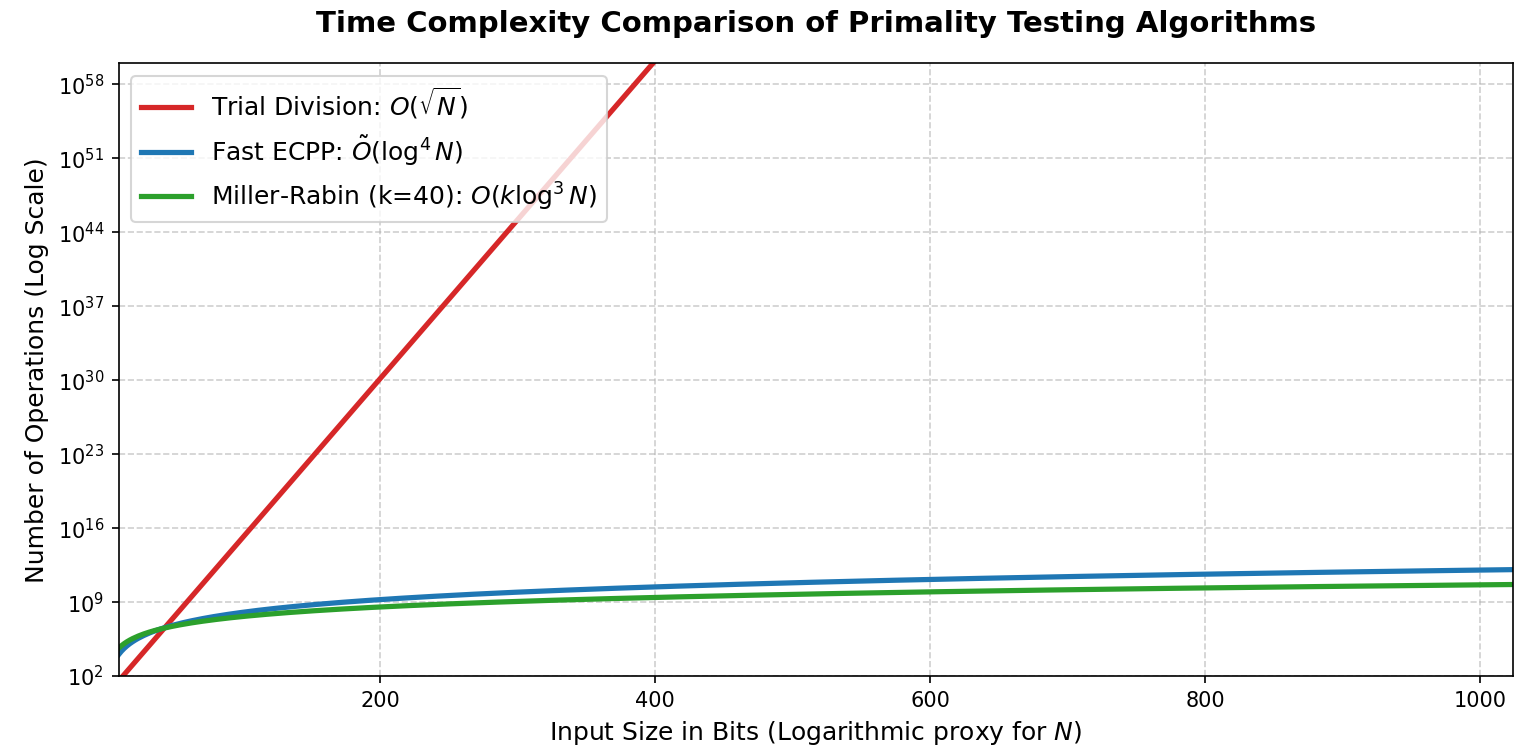}
    \caption{Theoretical operational scaling and complexity boundaries of primality testing frameworks (analytical operation count plotted as a function of candidate bit length), illustrating the combinatorial explosion of Trial Division $O(\sqrt{N})$ against polynomial-time alternatives such as Miller-Rabin and FastECPP.}
    \label{fig:time_complexity}
\end{figure}

Notably, across bounded numerical intervals, the Miller-Rabin test can transcend standard probabilistic behavior to operate deterministically without error. While the general algorithm exhibits an asymptotic error probability bounded by $(1/4)^k$ as surveyed by~\cite{ref1}, number-theoretic analyses demonstrate that evaluating candidates against small, precomputed sets of fixed prime bases yields an exact deterministic test below established thresholds (such as 64-bit integers tested against the first 12 prime bases, as contextualized in foundational prime testing analyses~\cite{ref41}). Under these bounded conditions, it functions with absolute mathematical certainty and negligible computational latency. In empirical benchmarks, the Miller-Rabin algorithm exhibits runtime performance substantially superior to both Trial Division and Solovay-Strassen, requiring only microseconds for word-scale inputs where exhaustive trial division degrades significantly~\cite{ref1}, reflecting the divergence in analytical operation counts illustrated in Figure~\ref{fig:time_complexity}.

To augment security when processing large integers exceeding 64 bits, cryptographers often combine multiple testing methods into a heuristic hybrid process. Prominent among them is the Baillie-PSW method, which can be briefly understood as a combination of the Miller-Rabin algorithm (typically choosing base $a=2$) and the Lucas-Sequence test. The process of evaluating an integer $N$ takes place through three consecutive steps. First, $N$ undergoes trial division against small prime numbers to quickly eliminate simple cases, next $N$ undergoes a base-2 Miller-Rabin testing round; finally is the evaluation step using Lucas Sequences. If $N$ passes all three of these challenges, it will be recorded as a "probable prime" (or a BPSW pseudoprime). This method has extremely high practical reliability. To date, no composite number less than $2^{64}$ has been able to successfully "pass through" Baillie-PSW, and no one has found any general counterexamples existing in practice~\cite{ref1}.

\vspace{1em}
\noindent \textbf{Table \ref{tab:primality_tests}} summarizes the time complexities, error probabilities, and cryptographic viability of these frameworks.

\begin{table}[htbp]
\centering
\caption{Comparison of Cryptographic Primality Testing Algorithms}
\vspace{0.3cm}
\renewcommand{\arraystretch}{1.3}
\begin{tabularx}{\textwidth}{>{\hsize=.15\hsize}X >{\hsize=.15\hsize}X >{\hsize=.15\hsize}X >{\hsize=.25\hsize}X >{\hsize=.3\hsize}X}
\toprule
\textbf{Algorithm Framework} & \textbf{Time Complexity} & \textbf{Error Probability} & \textbf{Cryptographic Advantages} & \textbf{Vulnerabilities / Drawbacks} \\
\midrule
\textbf{Fermat Test} & $O(k \log^3 N)$ & 1 for Carmichael numbers & Simple baseline implementation & Utterly defeated by Carmichael numbers~\cite{ref1}. \\
\textbf{Solovay-Strassen} & $O(k \log^3 N)$ & $< (1/2)^k$ & Resolves Carmichael flaw & High overhead computing Jacobi symbols~\cite{ref1}. \\
\textbf{Miller-Rabin} & $O(k \log^3 N)$ & $< (1/4)^k$ & Extremely fast, tunable certainty & Requires multiple bases for absolute security~\cite{ref1}. \\
\textbf{Baillie-PSW} & $O(k \log^3 N)$ & 0 for $N < 2^{64}$ & No known pseudoprimes exist & Heuristic; lacks formal mathematical proof for all $N$~\cite{ref1}. \\
\bottomrule
\end{tabularx}
\label{tab:primality_tests}
\end{table}

\subsection{Las-Vegas Algorithms and Elliptic Curve Primality Proving (ECPP)}

While Monte-Carlo primality tests provide excellent efficiency for routine cryptographic prime generation, applications requiring a mathematically conclusive proof of primality may instead employ primality-proving procedures. In the taxonomy adopted by AbuDaqa et al., randomized implementations of Proth, Lucas, and Pocklington tests are classified as Las-Vegas primality tests: a successful primality witness yields an exact conclusion, while the time required to obtain such a witness may depend on randomized choices~\cite{ref1}.

Proth's test, first proposed in 1878, is specialized to Proth numbers $N = k \cdot 2^n + 1$, where $k$ is a positive odd integer satisfying $k < 2^n$. A candidate base $a$ may be selected such that the Jacobi symbol $(a/N) = -1$. If
$$a^{(N-1)/2} \equiv -1 \pmod N$$
then $N$ is proven prime. When $N$ is prime, exactly half of the nonzero residue classes modulo $N$ satisfy the corresponding quadratic non-residue condition, allowing a suitable witness to be found efficiently~\cite{ref1}.

For generic integers, Lucas- and Pocklington-type criteria provide classical $N-1$-based approaches to primality proving. The Lucas criterion establishes the primality of an odd integer $N$ if there exists an integer $a$ such that
$$a^{N-1} \equiv 1 \pmod N$$ and $$a^{(N-1)/q_i} \not\equiv 1 \pmod N$$
for every prime divisor $q_i$ of $N-1$. In a simplified form of Pocklington's criterion, it is sufficient to know a prime divisor $q \mid N-1$ satisfying $q > \sqrt{N} - 1$, together with the required modular-exponentiation and gcd conditions. More generally, Pocklington's theorem can operate with a sufficiently large completely factored portion of $N-1$~\cite{ref1}.

The principal limitation of these $N-1$-based methods is their dependence on factorization information about $N-1$. The Lucas criterion requires the prime divisors of $N-1$ to be known, whereas Pocklington's criterion reduces this requirement to a sufficiently large completely factored portion of $N-1$. For arbitrary cryptographic-size candidates, obtaining such factorization information may introduce substantial computational overhead. Consequently, these methods are particularly attractive when primes are constructed so that suitable factors of $N-1$ are known in advance.

To avoid this dependence on the factorization of $N-1$, general-purpose primality proving can instead employ Elliptic Curve Primality Proving (ECPP), which does not require the candidate integer to have a special algebraic form and produces independently verifiable primality certificates. Elliptic-curve-based primality testing was first developed by Goldwasser and Kilian, while Atkin subsequently introduced a more practical approach based on Complex Multiplication (CM), which was implemented and extensively developed with Morain into the algorithm now commonly known as ECPP~\cite{ref19}.

ECPP operates with elliptic curves modulo the candidate integer $N$. Under the hypothesis that $N$ is prime, such a curve corresponds to an elliptic curve over the finite field $\mathbb{F}_N$. For an elliptic curve $E$ defined over a finite field $\mathbb{F}_p$, Hasse's theorem establishes the fundamental size constraint used in elliptic-curve primality criteria~\cite{ref18}:
$$p + 1 - 2\sqrt{p} \leq \#E(\mathbb{F}_p) \leq p + 1 + 2\sqrt{p}$$
Using CM theory, ECPP searches for a discriminant $D$ for which the candidate $N$ satisfies an appropriate norm equation and derives candidate elliptic-curve orders $m$ near $N+1$. Rather than requiring $m$ itself to be small, the algorithm seeks an order having a favorable partial factorization $m = F \cdot q$, where $F$ is completely factored and $q$ is a large probable-prime factor~\cite{ref19}. In a simplified single-factor formulation of the elliptic-curve primality criterion, $q$ must satisfy $q > (N^{1/4} + 1)^2$. For an appropriately constructed point $P$, the algorithm verifies that $mP = \mathcal{O}$, while the point $\frac{m}{q}P = (x_q:y_q:z_q)$ satisfies the required coprimality condition $\gcd(z_q, N) = 1$. Once these conditions hold and $q$ is itself proven prime, the primality of $N$ follows from the elliptic-curve criterion~\cite{ref19}.

The proof is therefore reduced recursively from $N = N_0$ to a smaller probable prime $N_1 = q$, and the same procedure is repeated to construct a decreasing chain $N_0 > N_1 > \dots > N_k$. The recursion terminates when the remaining candidate is sufficiently small to be verified directly by a simple deterministic method. The intermediate curve parameters, point coordinates, curve orders, and relevant factors form a primality certificate that can subsequently be checked by an independent verifier without repeating the computationally expensive search for suitable curves~\cite{ref19}. The running time of conventional ECPP has not been rigorously established; however, heuristic analysis gives an expected complexity of approximately $\tilde{O}((\log N)^5)$ bit operations~\cite{ref16}. An asymptotically faster variant, commonly referred to as fastECPP and attributed to J. O. Shallit in Morain's analysis, has a heuristic complexity of $\tilde{O}((\log N)^4)$, with practical implementations capable of handling integers containing several thousand decimal digits~\cite{ref16, ref_ecpp_record}. Furthermore, hybrid formulations combining one round of ECPP with an AKS iteration have been established by~\cite{ref17} to achieve deterministic polynomial-time verification. ECPP is therefore regarded as one of the fastest practical general-purpose methods for obtaining rigorous primality proofs for large integers, while retaining the important advantage that the resulting proof certificate can be independently verified.

\begin{figure}[htbp]
    \centering
    \includegraphics[width=0.75\textwidth]{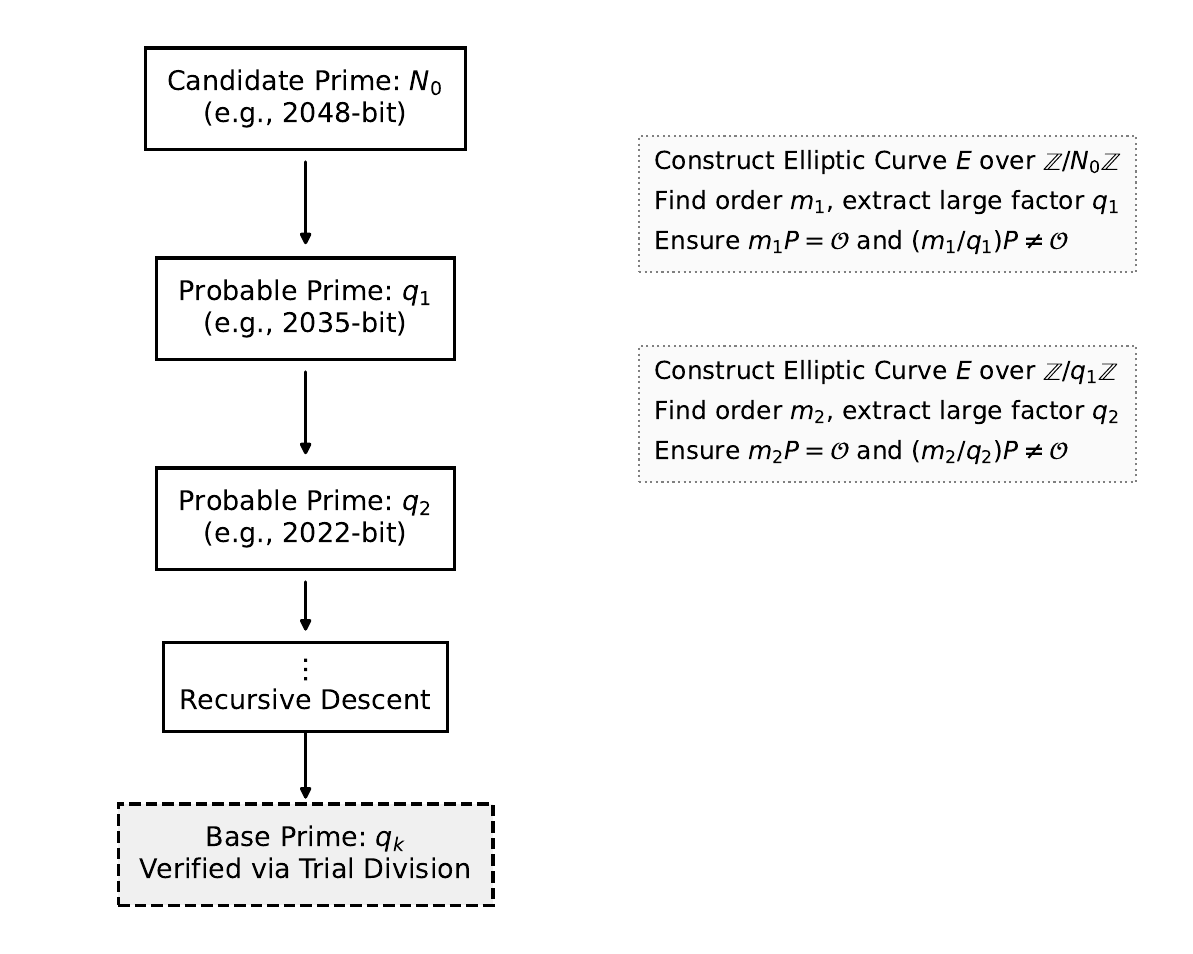}
    \caption{Recursive ECPP proof chain. Starting from the candidate $N_0$, each step reduces the proof to a smaller probable prime $q_i$ until a small prime is verified directly by trial division. Bit lengths are illustrative only.}
    \label{fig:ecpp_tree}
\end{figure}

\section{Algorithmic Sieving and Spatial Complexity Optimization}

To efficiently locate a prime number within a massive number space, the search algorithm needs to work closely with primality testing and proving tools, such as Miller-Rabin or ECPP. This algorithm plays the decisive role in determining the sequence of candidate numbers fed into processing. If one applies only a "brute-force" randomized search method on 2048-bit numbers and immediately runs probabilistic tests, the system will be extremely wasteful of computational resources and expensive modular exponentiations on easily recognizable composite numbers. Therefore, to optimize processing speed, the search space must be pre-filtered using sieving algorithms.

The classical Sieve of Eratosthenes is a simple, highly effective method where the algorithm iterates from 2 and sequentially marks all multiples of each prime as composite~\cite{ref22}. This method achieves an asymptotic time complexity of $O(N \log \log N)$~\cite{ref22}. However, when sieving over broad intervals, spatial complexity increases linearly as $O(N)$, which easily triggers Memory Limit Exceeded (MLE) faults on modern architectures when $N$ exceeds standard memory capacities.

The solution to this limitation is the Segmented Sieve of Eratosthenes, foundational formulations of which were developed by~\cite{ref22} and comprehensively surveyed by~\cite{ref24}. Instead of allocating an entire monolithic array in RAM, the algorithm partitions the sieving interval into contiguous blocks of width $\Delta$. In our hardware-aware implementation, we adapt this classical segmentation principle by sizing $\Delta$ specifically to reside entirely within the modern CPU L1 data cache (e.g., 32KB), thereby minimizing cache-miss penalties during bit-marking sweeps. In our cryptographic architecture, this segmented engine serves dual operational roles: (1) constructing a precomputed prime table up to a moderate bound $\sqrt{R}$ to amortize trial division costs, and (2) sieving candidate offsets within parameterized arithmetic progressions of the form $2RF + 1$ as utilized in constructive prime generation frameworks~\cite{ref41}, leveraging the wheel and interval sieving mechanics described in~\cite{ref22, ref24}.

\begin{figure}[htbp]
    \centering
    \includegraphics[width=0.9\textwidth]{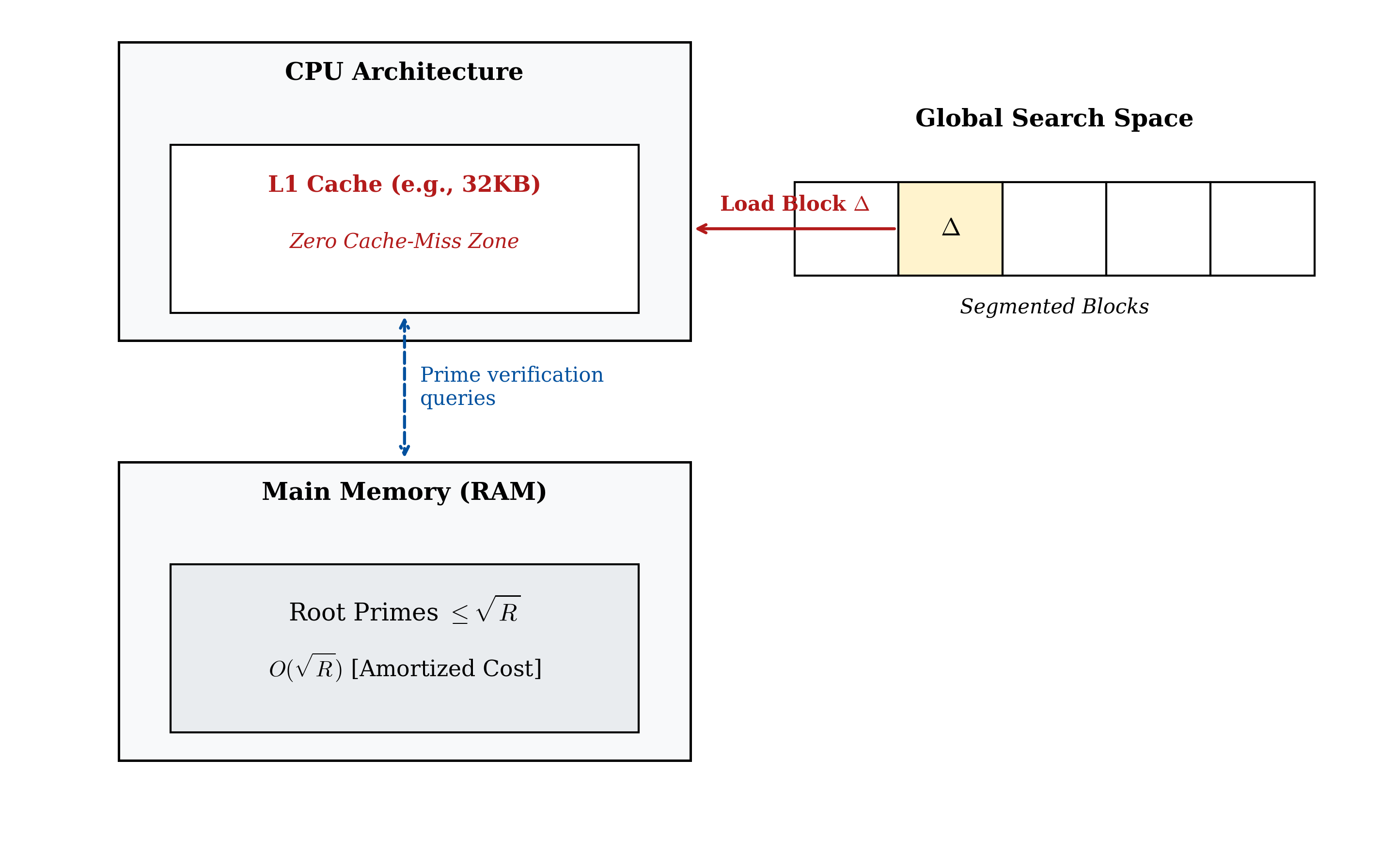}
    \caption{Hardware-aware memory architecture for the Segmented Sieve of Eratosthenes. The search space is dynamically partitioned into intervals of size $\Delta$, tuned to fit within the L1 Cache (e.g., 32KB) to eliminate cache-miss penalties.}
    \label{fig:l1_cache}
\end{figure}

By retaining only the boolean array for the active segment alongside the precomputed base primes up to $\sqrt{R}$, the spatial footprint drops precipitously from $O(N)$ down to $O(\sqrt{R} + \Delta)$ bits (see Figure~\ref{fig:l1_cache})~\cite{ref22, ref24}. Nevertheless, practical deployment reveals the \textit{initialization overhead paradox}: for isolated, high-entropy single-candidate queries at cryptographic scales ($N \ge 1024$ bits), initializing a broad sieve array introduces an unnecessary latency penalty.

Specifically, when generating or verifying isolated cryptographic candidates, generating an exhaustive prime base up to $\sqrt{R}$ represents extreme one-time initialization costs that cannot be amortized. Conversely, directly routing these isolated candidates to the Montgomery-accelerated Miller-Rabin testing pipeline completes the evaluation in a fraction of the time. 

For this reason, a highly adaptive cryptographic framework must autonomously evaluate the topological "scale" and density of the search domain prior to execution. The system acts as a heuristic dispatcher: broad candidate clusters and precomputed prime bases utilize the cache-optimized Segmented Sieve, whereas isolated point queries immediately bypass the sieve stage and proceed directly to the arbitrary-precision Montgomery-accelerated testing pipeline to prevent initial overhead from paralyzing the system.

\section{Bridging Number Theory and Machine Learning}

\subsection{Limitations of Continuous Optimization in Modular Arithmetic}

In applied computer science, the explosion in data scale and the number of parameters has turned continuous optimization methods, represented by Gradient Descent and its variants, into core tools for minimizing complex functions. Seeking to connect these continuous optimization algorithms with discrete algorithmic number theory opens a promising new analytical direction, but along with it come no few     challenges~\cite{ref_dl_textbook}.

Fundamentally, the Gradient Descent algorithm searches for minima (global or local) by computing the partial derivatives of a continuous objective (or loss) function $J(\theta)$ with respect to the input parameters $\theta$. Through each iteration, the model updates parameters in the direction of the steepest negative gradient:
$$\theta_{t+1}=\theta_t-\eta\cdot\nabla_\theta J(\theta)$$
where $\eta$ serves as the learning rate. Depending on the scale of data used to compute the gradient at each step, this method is divided into well-known common variants such as Batch Gradient Descent (computed over the entire dataset), Stochastic Gradient Descent (SGD) (updated per single sample, resulting in large oscillation amplitudes), and Mini-batch Gradient Descent (the standard in practice, helping to balance stability and computational cost)~\cite{ref_dl_textbook}.

To improve the ability to move across non-convex loss surfaces, advanced algorithms integrating "momentum" were developed, helping the model quickly overcome ravines or saddle points. Among them, Vanilla Momentum accumulates update vectors from previous steps to eliminate redundant oscillations:
$$v_t=\gamma v_{t-1}+\eta\nabla_\theta J(\theta)$$
Meanwhile, Nesterov Accelerated Gradient (NAG) goes one step ahead by applying momentum prior to computing the gradient, allowing the model to proactively adjust its path in a predictive manner~\cite{ref_dl_textbook}:
$$v_t=\gamma v_{t-1}+\eta\nabla_\theta J(\theta-\gamma v_{t-1})$$

Adaptive learning rate methods help dynamically scale parameter update step sizes based on historical gradient characteristics. Among them, Adagrad divides the learning rate by the square root of the accumulated sum of squared gradients; although highly effective with sparse data, Adagrad suffers from the drawback that the learning rate decays monotonically over time. To thoroughly address this, Adadelta and RMSprop substituted the infinite cumulative sum with an exponentially decaying moving average of past squared gradients:
$$E[g^2]_t=\gamma E[g^2]_{t-1}+(1-\gamma)g_t^2$$
Considered the apex within this group of methods, the Adam (Adaptive Moment Estimation) algorithm manages both the first moment $m_t$ (mean) and the second moment $v_t$ (variance) of the gradients in parallel, while incorporating bias correction to yield extremely stable weight update steps~\cite{ref_dl_textbook}:
$$\theta_{t+1}=\theta_t-\frac{\eta}{\sqrt{\hat{v}_t}+\epsilon}\hat{m}_t$$
Nadam further enhances Adam by incorporating Nesterov momentum directly into the first moment calculation~\cite{ref_dl_textbook}.

\begin{figure}[htbp]
    \centering
    \includegraphics[width=1\textwidth]{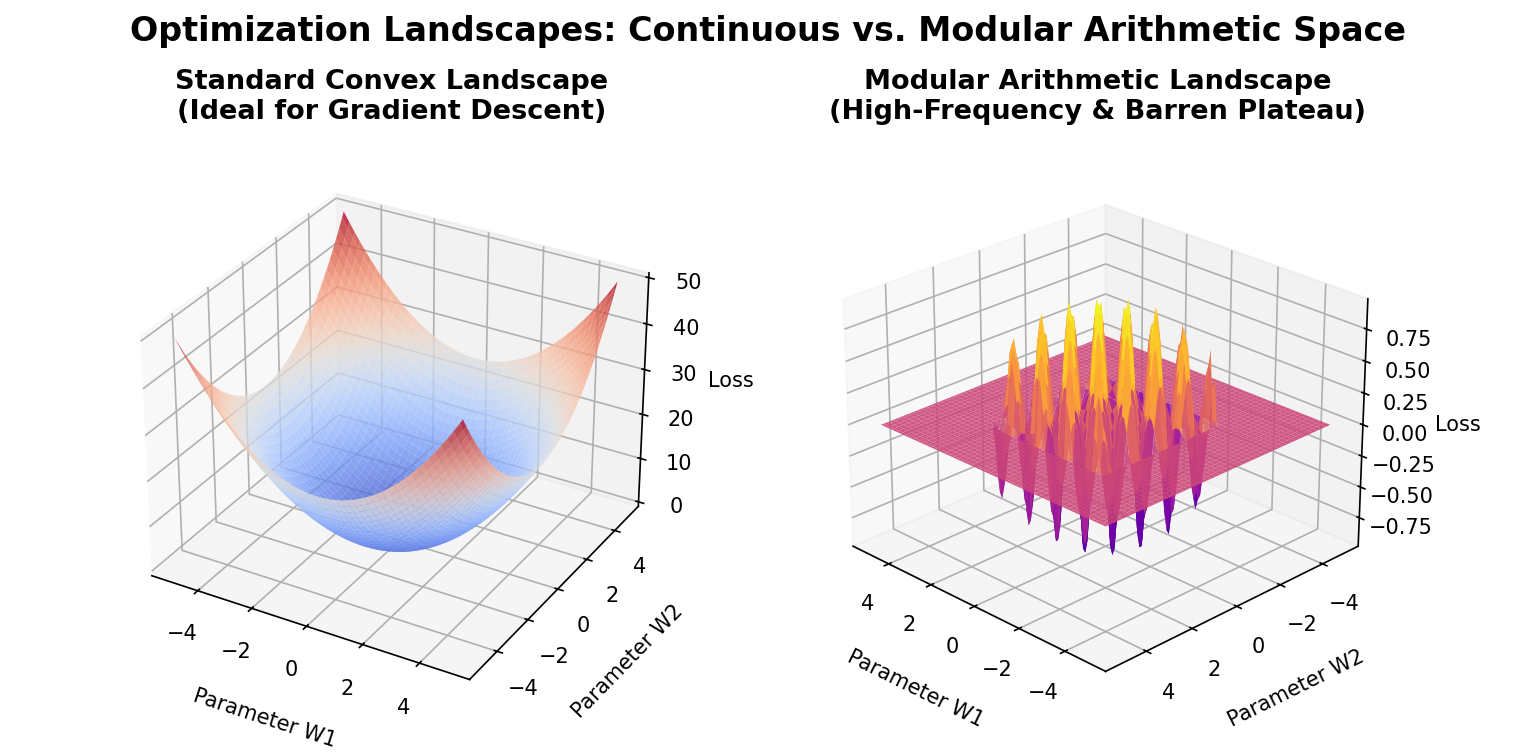}
    \caption{Comparison of Optimization Landscapes: Continuous Standard Convex vs. Modular Arithmetic Space. The high-frequency oscillation in modular arithmetic creates the "Barren Plateau" phenomenon where gradient information is effectively lost.}
    \label{fig:barren_plateau}
\end{figure}

Although Adam, RMSprop, or Nadam possess profound capability in training deep neural networks and optimizing over continuous vector spaces, applying continuous Gradient Descent directly to discrete arithmetic problems encounters insurmountable mathematical barriers~\cite{ref7}. ~\cite{ref7} proved that learning modular arithmetic mappings (such as $x \mapsto ax \pmod p$) or high-frequency periodic functions induces the severe "barren plateau" phenomenon: the variance of the gradient w.r.t. the hypothesis space vanishes asymptotically as $\tilde{O}(p^{-1/2})$, depriving gradient-based optimizers of directional information. In optimization landscapes, this rapid variance decay creates vast flat regions where gradient updates collapse toward zero, causing continuous optimizers to lose direction entirely and fail to optimize modular targets \textbf{(as visualized in Figure \ref{fig:barren_plateau})}.

Consequently, standard continuous gradient-based optimization algorithms are completely 'disoriented' when operating in modular arithmetic spaces, making the application of gradient descent methods to directly predict primality or factorization impossible.

\subsection{Discrete Optimization and ML-Guided Witness Selection}

Because direct continuous gradient optimization collapses in high-frequency modular landscapes, optimization must be elevated to meta-level pipeline calibration and discrete decision models. In our framework, Bayesian Optimization serves as a principled paradigm for tuning black-box operational hyperparameters—such as empirically balancing trial-division thresholds alongside theoretical analytical optima~\cite{ref41} and configuring multiple-precision pipeline parameters—by navigating exploration-exploitation trade-offs via probabilistic surrogate models and entropy-search acquisition strategies~\cite{ref31, ref5}. Furthermore, drawing a methodological analogy from systems research where reinforcement learning and policy gradient principles~\cite{ref34} are deployed to optimize software cache allocations, data retention, and telemetry-driven buffer policies (as exemplified by Alabed~\cite{ref37}), our framework models the dynamic management of scratchpad memory pools and batch scheduling buffers as an adaptive resource-allocation process.

At the algorithmic level, we investigate an exploratory Proof-of-Concept (PoC) machine learning pipeline for intelligent witness selection in the Miller-Rabin test. While random bases eliminate ordinary composites with probability $\ge 3/4$ per round~\cite{ref14}, adversarial candidates and strong pseudoprimes can survive specific standard bases. To accelerate early termination without introducing latency bottlenecks, our framework extracts lightweight structural features from candidate $N$ (e.g., bitwise topology, low-degree residue vectors, and localized Hamming weights) in linear time $\tilde{O}(\text{len}(N))$.

A pre-trained Random Forest model ranks candidate witness bases to maximize the probability of triggering early composite rejection on the first trial. Crucially, to prevent feature extraction from creating an arbitrary-precision bottleneck, all feature operations are hardware-accelerated via native word-level primitives—primarily bit-shifts, trailing zero counts, and modular arithmetic restricted strictly to small word-sized primes. These design choices constrain the total feature preparation cost to strictly linear time $\tilde{O}(\text{len}(N))$, which is orders of magnitude cheaper than the super-linear $O(\text{len}(N)^2)$ or $O(\text{len}(N)^3)$ cost of a full arbitrary-precision modular exponentiation.

Deploying this lightweight ML-driven witness selector equips the framework with a specialized defense against adversarial candidates and highly structured pseudoprimes. In standard execution, candidate evaluation initiates with the fixed base $a = 2$, which filters out the overwhelming majority of arbitrary composite numbers at minimal cost. However, in adversarial cryptographic environments, malicious inputs or pathological pseudoprimes (such as engineered Carmichael or Arnault-type structures) are deliberately crafted to pass base-2 evaluations.

For candidates surviving this initial sieve, the ML-driven witness selector is activated: rather than proceeding through arbitrary random bases that may require extensive iterations to hit a non-trivial witness, the pre-trained classifier evaluates localized bitwise and residue features to prioritize witness bases with the highest statistical likelihood of triggering early rejection on structured composites. If the candidate is indeed prime, the pipeline completes the required round budget; however, for complex adversarial composites, early termination is achieved on the initial non-trivial trial, circumventing redundant high-precision exponentiations and saving computational overhead.

\begin{figure}[htbp]
    \centering
    \includegraphics[width=1\textwidth]{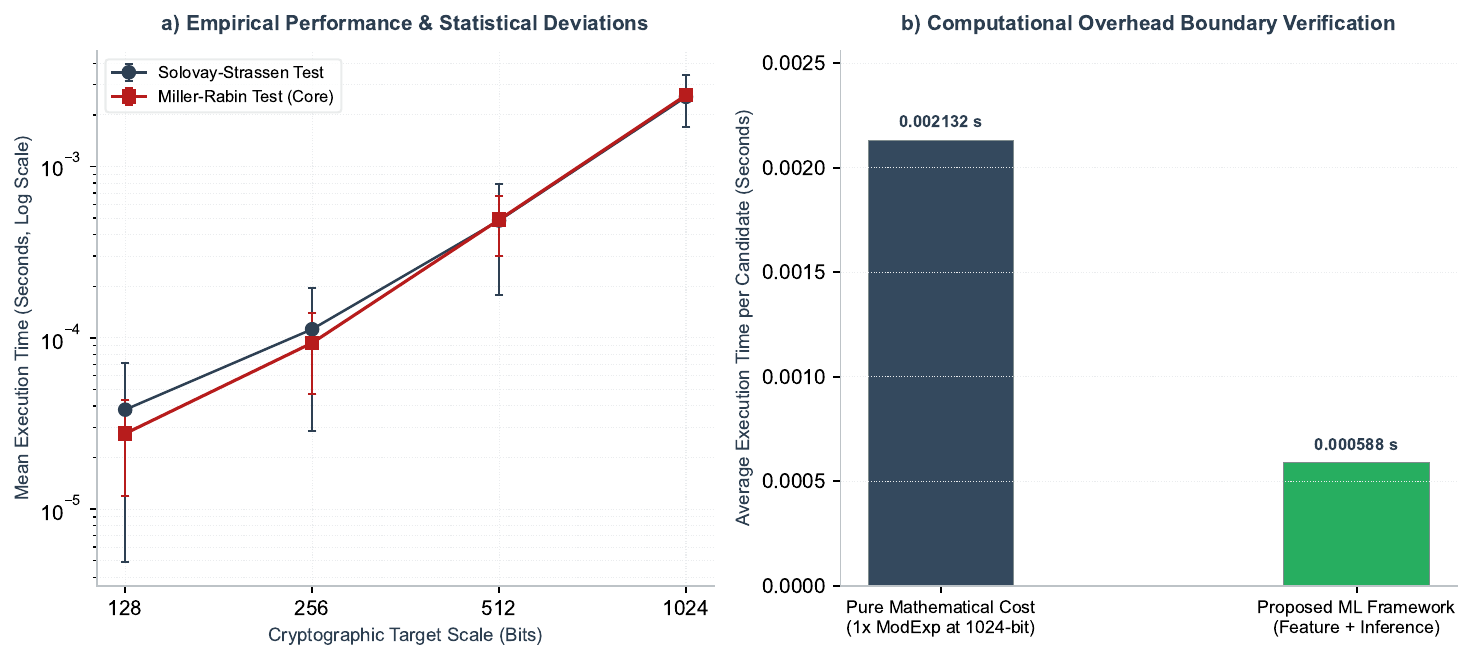}
    \caption{Empirical verification of the cryptographic optimization framework: (a) Runtime performance under scaling cryptographic constraints ($N=500$), contrasting the operational variance and lower mean execution time of the Miller-Rabin test against the Solovay-Strassen baseline; (b) Computational overhead verification demonstrating compliance with the core architectural boundary condition, where the proposed ML infrastructure remains significantly cheaper than a single hardware modular exponentiation.}
    \label{fig:experimental_framework_benchmarks}
\end{figure}

To convincingly validate the theoretical claims and ensure the ML model can be effectively deployed in practice, we conducted extensive performance benchmarking. All empirical benchmarks were conducted in a Python 3.10 environment on a single CPU core of an Intel Core i7-12700H @ 2.30 GHz host (Asus TUF Gaming F15 architecture) with 16 GB RAM, where measurements were captured using monotonic high-resolution hardware timers (\texttt{time.perf\_counter()}) and averaged over $500$ independent samples. As depicted in Figure \ref{fig:experimental_framework_benchmarks}(a), the execution time of classical algorithms strictly adheres to theoretical asymptotic complexity. Among them, the core Miller-Rabin algorithm consistently achieves superior performance and operates more stably (with lower operational variance) compared to the Solovay-Strassen baseline due to the complete elimination of costly Jacobi symbol evaluations.

In particular, Figure \ref{fig:experimental_framework_benchmarks}(b) provides decisive evidence for perfectly meeting the computational overhead constraints specified in Table \ref{tab:optimization}. Specifically, the average execution time of the entire supervised machine learning pipeline, including both advanced non-linear feature extraction and inference via a lightweight Random Forest model, registers at a mere $0.000588$ seconds per candidate. When compared to the average cost of $0.002132$ seconds required to perform a single traditional 1024-bit modular exponentiation ($\text{pow}(a, d, n)$), the intelligent witness selector consumes only $27.58\%$ of the computational budget. This empirical verification result guarantees that the framework significantly accelerates the witness discovery process without introducing any software latency bottleneck, reinforcing its high applicability in cryptographic architectures requiring ultra-fast response times.

\vspace{1em}
\noindent \textbf{Table \ref{tab:optimization}} outlines the various optimization paradigms and their cryptographic applications.

\begin{table}[htbp]
\centering
\caption{Optimization Paradigms in Algorithmic Number Theory}
\vspace{0.3cm}
\renewcommand{\arraystretch}{1.3}
\begin{tabularx}{\textwidth}{>{\hsize=.2\hsize}X >{\hsize=.2\hsize}X >{\hsize=.25\hsize}X >{\hsize=.35\hsize}X}
\toprule
\textbf{Optimization Paradigm} & \textbf{Cryptographic Application} & \textbf{Mechanism of Action} & \textbf{Constraint / Limitation} \\
\midrule
\textbf{Gradient Descent (Adam, RMSprop)} & Neural Cryptanalysis & Continuous weight updates & Fails due to barren plateaus in high-frequency modular arithmetic~\cite{ref7}. \\
\textbf{Bayesian Optimization} & Search Grid Tuning & Surrogate probabilistic models & Computationally expensive scaling GP models for massive sample budgets. \\
\textbf{Policy Gradients}~\citep{ref34} & Dynamic Cache \& Buffer Tuning & Reinforcement over execution telemetry & Requires continuous statistical sampling and validation of state transitions~\citep{ref37}. \\
\textbf{Supervised Machine Learning} & Intelligent Witness Selection & Predictive targeting of prime bases & Classifier overhead must remain smaller than modular exponentiation cost. \\
\bottomrule
\end{tabularx}
\label{tab:optimization}
\end{table}

\section{Cryptographic Security of Biased Search Distributions}

Heuristic candidate generation and constrained prime search (e.g., searching within arithmetic progressions $2RF + 1$ or incrementing by odd steps) inherently induce a non-uniform distribution over the prime space~\cite{ref41}. Historically, any distributional bias was viewed with suspicion due to concerns that adversaries could focus factorization or discrete logarithm algorithms on higher-probability density regions.

This foundational concern was resolved by~\cite{ref47}, who established that prime generation sources with polynomially bounded bias maintain computational security equivalent to perfectly uniform sources under generic attacks. Formally, any algorithm capable of exploiting a polynomial distributional bias to accelerate the factorization of an RSA modulus $N = pq$ can be polynomial-time transformed into an algorithm that breaks uniformly distributed instances with comparable asymptotic efficiency~\cite{ref47}.

However, recent cryptanalytic breakthroughs by~\cite{ref45} reveal a crucial distinction between \textit{static distributional bias} and \textit{sequential temporal correlation}. While static polynomial bias is secure under Mih\u{a}ilescu's theorem, algorithmic shortcuts such as the bitwise overlapping candidate search introduce temporal correlations that recurrent deep learning architectures (CNN-LSTM) can exploit to predict future prime intervals with statistical significance exceeding 10 standard deviations, even when underpinned by quantum entropy~\cite{ref45}. 

Consequently, to ensure absolute cryptographic soundness, our framework strictly enforces a \textit{non-overlapping extraction mode} for all physical entropy inputs, guaranteeing independent candidate streams while leveraging Mih\u{a}ilescu's bounded-bias theorem to validate search-step optimizations.

\subsection{Quantum Entropy Integration in Candidate Generation}
To insulate the architecture against seed-reconstruction and state-exhaustion attacks inherent to deterministic PRNGs, the pipeline anchors its initial entropy in physical Quantum Random Number Generators (QRNGs)~\cite{ref46}. By measuring the quadrature fluctuations of the optical vacuum state via balanced homodyne detection, the system acquires genuine physical entropy with an established min-entropy rate exceeding 8.39 bits per 16-bit sample~\cite{ref45, ref46}. This quantum entropy guarantees that the starting coordinates across the arbitrary-precision search lattice remain strictly unpredictable, satisfying the physical prerequisite of Mih\u{a}ilescu's security theorem. 

\section{Synthesis of the Adaptive Framework}

\begin{figure}[htbp]
    \center
    \makebox[\textwidth][c]{\includegraphics[width=1.3\textwidth, keepaspectratio]{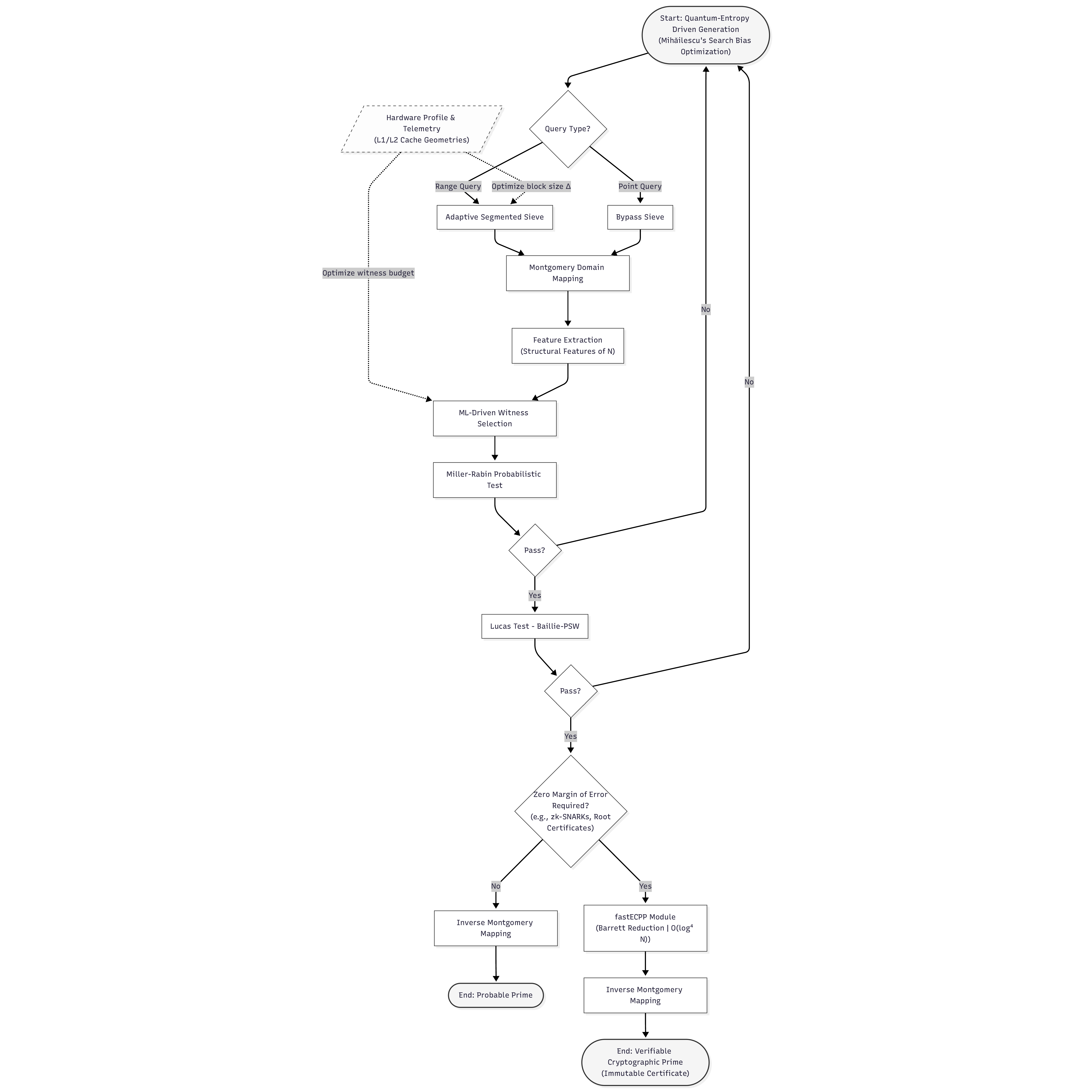}}
    \caption{Global Architectural Flowchart of the hybridized, adaptive primality testing framework. The diagram illustrates the quantum-anchored generation phase and the dynamic selection of algorithmic pathways based on input topology and hardware constraints.}
    \label{fig:flowchart}
\end{figure}

By closely combining the mathematical proofs of bounded bias (Section 6) and physical quantum entropy protocols (Section 6.1), this adaptive framework establishes a solid bridge between classical number theory and modern optimization paradigms. \textbf{As illustrated in Figure \ref{fig:flowchart}}, the system no longer operates as a rigid monolithic processing block; instead, the entire pipeline is secured right from the initialization step thanks to an unconditionally secure quantum random number source. Grounded on this perfect entropy foundation, the system can freely and flexibly restructure its execution pipelines, from optimizing sieving algorithms to integrating machine learning (ML) heuristic filters, aimed at maximizing search throughput without compromising cryptographic security.

The candidate generation process is initialized using physical noise sources rich in entropy, wherein quantum entropy fundamentally outperforms classical noise to ensure true, unconditional unpredictability~\cite{ref45, ref46}. Grounded on Mih\u{a}ilescu's theorem, the system can apply search steps with polynomially bounded biases to maximally accelerate the identification of potential candidates while fully maintaining the original cryptographic security~\cite{ref47}.  

The first filtering layer employs an adaptive Segmented Sieve of Eratosthenes algorithm. The system automatically analyzes the host CPU hardware architecture, including L1 and L2 cache sizes, to tune the block size parameter $\Delta$, keeping the data entirely within the cache and completely eliminating latency caused by cache misses~\cite{ref24}. In particular, when encountering isolated point queries that evaluate a single large integer, the system identifies the computational inefficiency, proactively bypasses the sieving stage to avert wasting the $O(\sqrt{R})$ array initialization overhead, and routes the candidate directly to the probabilistic testing pipeline.  

Candidates surviving the preliminary filtering stage enter the Miller-Rabin testing module. Here, arbitrary precision large integer array structures are directly mapped into the Montgomery domain, replacing costly division operations with extremely fast binary bit shifts throughout the modular exponentiation process. Instead of randomly selecting testing bases, the system employs a pretrained supervised machine learning model. This model analyzes the structural features of the integer $N$ to rank and accurately predict the bases with the highest probability of exposing compositeness as witnesses. Consequently, Miller-Rabin can rapidly eliminate pseudoprimes after just the first one or two trials. Surviving candidates are then further verified via Lucas sequences under the Baillie-PSW paradigm to completely eliminate all known classes of strong pseudoprimes and Carmichael numbers.  

Finally, for high-assurance cryptographic infrastructures requiring zero margin of error (such as zk-SNARK trusted setups, root certificate authorities, and verifiable key registries), probabilistic testing is insufficient. The architecture seamlessly routes validated probable primes to the fastECPP module~\cite{ref16}. At this stage, Barrett reduction is strategically deployed to accelerate the extensive modular arithmetic required during the downrun phase—most notably the computationally dominant root-finding of Hilbert and Weber class polynomials $H_D(X) \pmod{N_i}$ via Cantor-Zassenhaus splitting~\cite{ref16, ref19}. Because the intermediate candidate modulus $N_i$ remains invariant across thousands of polynomial coefficient reductions within each downrun level, precomputing Barrett's reciprocal invariant $\mu = \lfloor b^{2k}/N_i \rfloor$ amortizes division overhead and avoids repetitive domain-transformation costs associated with Montgomery arithmetic across large polynomial arrays~\cite{ref2, ref10}. Executing in heuristic $\tilde{O}(\log^4 N)$ time, fastECPP constructs an independently verifiable hierarchical certificate of primality~\cite{ref16, ref19}. Because preceding filtering stages have eliminated non-prime candidates, the computational investment of ECPP is preserved exclusively for authentic cryptographic primes.

\section{Conclusion and Future Work}

This study establishes an adaptive primality testing framework, bridging classical number theory, discrete machine learning, hardware memory optimization, and quantum randomness sources. The system effectively addresses the performance bottlenecks of large-scale cryptographic key generation through an intelligent self-dispatching mechanism, combining an L1 cache-optimized Segmented Sieve with machine-learning-assisted witness selection for the Miller-Rabin test. Empirical results demonstrate that the feature extraction and inference pipeline consumes only $27.58\%$ of the runtime required for a single 1024-bit modular exponentiation, enabling early elimination of composite candidates without introducing software latency bottlenecks. Concurrently, anchoring candidate generation in quantum entropy alongside Mihăilescu's bounded-bias theorem preserves mathematical security up to deterministic verification via fastECPP.

In future work, the framework can be extended within the context of Post-Quantum Cryptography (PQC). A prospective direction involves identifying structured, Number Theoretic Transform (NTT)-friendly primes of the form $q \equiv 1 \pmod{2n}$ to support parameter selection for generalized lattice-based schemes. Additionally, transitioning the software prototype to high-performance systems languages (such as C or Rust) and evaluating hardware-accelerated implementations on FPGAs or ASICs will help assess practical throughput in specialized environments.

\bibliographystyle{unsrtnat}
\bibliography{references}

\end{document}